\documentclass[12pt,notitlepage,longbibliography]{article}
\usepackage{amsmath, mathrsfs, amssymb,amsfonts,amsthm,graphicx, epsf, dcolumn, yfonts, xcolor}
\usepackage[utf8]{inputenc}
\usepackage[super,sort&compress]{natbib}
\usepackage[normalem]{ulem}
\usepackage{authblk}
\usepackage[hyperfootnotes=true]{hyperref}
\usepackage{color}
\usepackage{float}
\usepackage{todonotes}
\usepackage{array}
\usepackage{tabularray}
\newcommand{\ket}[1]{\ensuremath{|{#1}\rangle}}

\pdfoutput=1

\begin{document}

\title{Classical gravity cannot mediate entanglement}

\author[1]{Chiara Marletto}
\affil[1]{Clarendon Laboratory, University of Oxford, Parks Road, Oxford OX1 3PU, United Kingdom}
\author[2]{Jonathan Oppenheim}
\affil[2]{Department of Physics and Astronomy, University College London, Gower Street, London WC1E 6BT, United Kingdom}
\author[1]{Vlatko Vedral}
\author[2]{Elizabeth Wilson}
\date{}
\maketitle
\vspace{-6em}
\renewcommand{\abstractname}{}
\begin{abstract}
In Nature, 646, 813 (2025), Aziz and Howl claim that classical (unquantised) gravity produces entanglement. We show that their model does not produce entanglement. Even if the model produced entanglement, it would be mediated by the quantised matter interaction, and not gravity. Hence entanglement mediated by gravity remains  an unambiguous witness of gravity's quantum features. 
\end{abstract}

Two recent proposals to detect quantum effects in gravity \cite{Bose,MV} use the so-called Gravitationally Mediated Entanglement (GME) between two massive probes to witness non-classical features in gravity. The proposals are based on the fact that if gravity can mediate entanglement between two quantum systems then it must itself have non-classical features, \cite{MV, MV-RMP, MAVE20}. The paper by Aziz and Howl (AH) \cite{Howl} presents a supposed counterexample -- a Hamiltonian $H_{int}$ (Eq. 4 in the AH paper\cite{Howl}) which is claimed to be capable of `producing' entanglement between two masses in superposition. The claim is incorrect and arises from an inconsistent coupling of the matter propagator, as well as from a necessary but insufficient criterion for what constitutes being ``responsible'' for entanglement.

The interaction Hamiltonian between the classical Newtonian potential $\Phi$ and the quantum scalar field $\hat{\phi}$ cannot generate entanglement as predicted in the article, because in the non-relativistic limit it acts as the sum  of terms $H_{int}\approx V_1+V_2$ on particle one and two, while in the subspace spanned by the states which form the initial superposition, the authors require the matter Hamiltonian to act as the sum $\hat H_0\approx \hat H_1+\hat H _2$ and therefore, the two taken together can only generate product unitaries which don't entangle.

That $H_{int}\approx V_1+V_2$, follows because the interaction term which leads to a 
direct coupling of  the two masses $\frac{\Phi}{c^2}(\nabla \hat\phi)^2$ 
is too small in the non-relativistic limit and is dropped in the actual calculation, leaving the ultra-local Hamiltonian, Eq. 41 of the Supplementary Information \cite{Howl}. 
Recall that an ultra-local Hamiltonian is one whose density depends only on local field operators and not derivatives and therefore can be written as $H_{int}\approx V_1+V_2$. The authors acknowledge this in the Discussion section of the Supplementary Information, but claim that it is the propagator in the quantum field $\hat{H}_0$ that contributes to the  entanglement  generating amplitude.

But adding the quantum Hamiltonian will not result in entanglement being produced either, because $\hat H_0\approx \hat H_1+\hat H _2$. This follows because the initial state of the separated particles is required to be in a product eigenstate of $\hat{H}_0$ (even though $\hat{H}_0$'s eigenstates are delocalised momentum states). As the authors  emphasise in the Supplementary Discussion, they require that $\hat H_0$ by itself produce only a global phase when acting on any of the states in the initial superposition  $|N\rangle_{1i}\otimes |N\rangle_{2j}$, (their Eq. 24) and therefore the initial product state $|\psi(0)\rangle=(|N\rangle_{1L}+|N\rangle_{1R})\otimes (|N\rangle_{2L}+|N\rangle_{2R})/2$ is an approximate eigenstate of $\hat H_0$. So in the  subspace spanned by the states that form the superposition, over the lifetime of the experiment $\hat H_0$ must also be a sum of terms $\hat H_0\approx \hat H_1+\hat H_2$ acting on each particle.  This is crucial for AH's argument that gravity is ``responsible'' for the entanglement, because their criterion is that $\hat H_0$ alone cannot create it. However, either $\hat H_0$ does not directly couple the two particles when acting on the initial state, in which case it gives rise to a product unitary; or it directly couples the two particles, in which case it alone will generate the entanglement, and no product eigenstate exists. It is not possible to have it both ways. Since $\hat H_0$ and $H_{int}$ are ultra-local in the subspace of the initial superposition, they lead to a product unitary when exponentiated and cannot produce entanglement.

How is the entanglement then obtained? By inconsistently dropping the momentum term $(\nabla\hat\phi)^2$ in $\hat H_0$ which propagates quantum information from particle 1 to particle 2. In Eq. 23, they drop the term, to obtain a time-dependent phase which only depends on the particle's mass, and not on $\textbf{x}$ and $\textbf{k}$. This is required in order that $\hat H_0$ not directly couple the two particles via the quantum field. But then when computing the propagator between the two particles, Eq. 63, it re-appears as $k^2$ in the denominator. Then when computing the amplitude via Eq. 66, they mix both the full propagator of Eq 63 and the ultra-local term from Eq. 23, and then integrate over all time (rather than just the time of the experiment) to obtain a $1/|x-y|$ interaction. That this is unphysical, can be seen from the fact that the propagator acquires a momentum given by the mass, but in the non-relativistic limit, this cannot come from either the particles (which are at rest), or the gravitational potential (which is stationary). Using the correct non-relativistic propagator instead gives a term proportional to $\delta(x-y)$ and the amplitude for the process is zero whenever the wavepackets of the two particles are not overlapping. 

We can now speculate about what the effect sought by AH could be. The propagator that AH hoped to trigger was that corresponding to the quantum matter field $\hat{\phi}$ of $\hat{H}_0$. In this case, gravity would, at best be a classical parameter modulating a direct interaction between matter terms. Direct matter-matter interactions are screened or suppressed in the original GME protocols \cite{Bose,MV}. Indeed one can always engineer models where a classical field manipulates interacting quantum systems into becoming entangled. This happens whenever an experimentalist — modelled classically — turns on a laser to generate entanglement, or when a classical clock triggers an interaction that harvests entanglement from the vacuum \cite{Reznik2003}. In such cases, the classical system does not mediate the entanglement. 

Even if the mechanism proposed by AH had worked, it would have been of this type: entanglement mediated by quantum matter triggered by a classical perturbation, not gravitationally mediated entanglement. Indeed  AH are essentially doing quantum field theory in a fixed external classical field. The criterion AH use to claim that gravity is ``responsible'' — that $\hat H_0$ alone does not produce entanglement until $H_{int}$ is ``switched on'' — is therefore clearly insufficient as it fails to distinguish their mechanism from any standard quantum interaction that is merely initiated by a classical system. We rectify their criterion in \cite{MOVW}. 

There, we also correct some misconceptions which have been amplified by the debate generated by this article. For example, the question of whether the ``classical Newtonian interaction'' can generate entanglement is an ambiguous one, which explains why it is so contested. When it arises from a classical theory of gravity it cannot generate entanglement. (See also \cite{MVC,O}). 
Note also that a truncation in the perturbative expansion can lead to the appearance of entanglement when there is none. While this is not the same interaction found by AH, a simple example will illustrate the point. Consider a product unitary $U=U_1\otimes U_2$, where $U_i$ acts on qubit $i$. As a product unitary, $U$ cannot generate entanglement, as it maps product states to product states: $U\ket{00}=(c\ket{0}+s\ket{1})(c\ket{0}+s\ket{1})$, with $c=\cos{\eta}$ and $s=\sin{\eta}$. For small $\eta$, we can expand $c$ and $s$ in Taylor series. At the second order in $\eta$, the state is still a product state; while at the fourth order it is entangled. So if we consider only specific terms of the expansion $U$ appears as an entangling operation, while overall it is not.

Finally, AH often state that the theorems about non-classicality GME witnesses rely on assumptions rooted in `standard quantum mechanics', rather than the field theory calculations they perform. This too is false, in that the theorems do not need to assume quantum theory's formalism or structure, only general information-theoretic axioms that are satisfied by all fundamental theories. In fact, the most general consistent form of quantum fields interacting with classical fields has been derived\cite{PQG,O2, LOW}, and explicitly shown to not generate entanglement via the classical field\cite{OW}. The mean field approach used in AH, while not mathematically consistent in the regime under consideration, has also been explicitly shown by Di\'osi to not generate entanglement when exactly solved\cite{D}.

In conclusion: the model proposed by AH, under their assumptions, cannot create entanglement. Models where quantum matter degrees of freedom interact directly can (obviously) lead to entanglement. Moreover, any c-numbered field (including classical gravity) that modulates the interaction does not mediate the entanglement. Detecting GME therefore remains an unambiguous witness of quantum effects in gravity.

Nonetheless, Aziz and Howl, and Weller-Davies in his commentary\cite{weller}, are right to stress the importance of understanding all possible routes by which two masses could become entangled. Because, as emphasised in the original experimental proposals\cite{Bose,MV}, the fact that classical fields do not mediate entanglement does not preclude the generation of entanglement through other quantum mechanisms, including exotic ones  such as via correlated environmental noise which appear in decoherence models~\cite{oppenheim2009fundamental,trillo2025diosi}. However, such effects, while interesting, are expected to be quantitatively different, in comparison to the entanglement predicted in  quantum theories of gravity. In either case, superpositions of massive particles remain as powerful quantum sensors to probe the quantum nature of spacetime.

\textbf{Acknowledgments:} This research was made possible by the generous support of the Gordon and Betty Moore Foundation, the Eutopia Foundation, and the Conjecture Institute. We thank Zach Weller-Davies, Richard Howl, Aditya Iyer, Lajos Di\'osi and Dan Carney for helpful discussions.

\bibliography{references}

\end{document}